\documentclass[aps,pra,groupedaddress,showpacs,floats]{revtex4}
\usepackage{amsmath,amsfonts,graphicx,mathrsfs,fancybox,epstopdf, setspace, latexsym}
  {\catcode`\|=\active \gdef|{\egroup\,\vrule\,\bgroup}}

\begin{document}
\title{Virtual dielectric waveguide mode description of a high-gain free-electron laser I: Theory}
\author{Erik Hemsing$^{\dagger}$, Avraham Gover$^{\ddagger}$ and James Rosenzweig$^{\dagger}$}
\affiliation{$\dagger$Particle Beam Physics Laboratory, Department of Physics and Astronomy\\
University of California Los Angeles, Los Angeles, CA 90095, USA}
\affiliation{$\ddagger$Faculty of Engineering, Department of Physical Electronics, Tel-Aviv University, Ramat-Aviv 69978, Tel-Aviv, Israel}
\date{\today}

\begin{abstract}
A set of mode-coupled excitation equations for the slowly-growing amplitudes of dielectric waveguide eigenmodes is derived as a description of the electromagnetic signal field of a high-gain free-electron laser, or FEL, including the effects of longitudinal space-charge. This approach of describing the field basis set has notable advantages for FEL analysis in providing an efficient characterization of eigenmodes, and in allowing a clear connection to free-space propagation of the input (seeding) and output radiation. The formulation describes the entire evolution of the radiation wave through the linear gain regime, prior to the onset of saturation, with arbitrary initial conditions. By virtue of the flexibility in the expansion basis, this technique can be used to find the direct coupling and amplification of a particular mode. A simple transformation converts the derived coupled differential excitation equations into a set of coupled algebraic equations and yields a matrix determinant equation for the FEL eigenmodes. A quadratic index medium is used as a model dielectric waveguide to obtain an expression for the predicted spot size of the dominant system eigenmode, in the approximation that it is a single gaussian mode.
\end{abstract}
\maketitle

\section{introduction}
The optical guiding of light in free-electron lasers (FELs) is a well-known phenomena that results during amplification when the coherent interaction between the source electron beam (e-beam) and the electromagnetic (EM) field introduces an inward curvature in the phase front of the light, refracting it back towards the lasing core of the e-beam\cite{Kroll, Moore, Scharlemann}. During the gain process the e-beam can behave like a guiding structure that suppresses diffraction, reducing transverse power losses and enhancing the EM field amplification.  For a sufficiently long interaction length, the guided EM field eventually dynamically settles into a propagating, self-similar eigenmode of the FEL system, which, although describing a system in which the field amplitude grows exponentially as a function of longitudinal coordinate $z$, has a fixed form for the transverse profile and stationary spot size\cite{Xie,Pinhasi}. 

Several different approaches currently exist for describing the FEL linear gain process. Analytic derivations of guided FEL eigenmodes have been performed previously by direct derivation of the eigenmode equations from the coupled Maxwell-Vlasov equations\cite{Xie,Xie2000a,Wu}; through solutions to the Maxwell equations using Green's functions\cite{Saldin}; or through expansion of the FEL signal field in terms of eigenmodes of a hollow, conducting-boundary waveguide\cite{Pinhasi}, eigenmodes of a step-index fiber\cite{Scharlemann}, and free-space paraxial waves\cite{Sprangle}.
Such analytic descriptions of the 3-D FEL equations, particularly those that exploit radiation mode expansions to find the FEL eigenmodes, have certain utility for providing physical insight into the character of the radiation. Since, in a high-gain FEL, the e-beam that propagates through the periodic undulator operates simultaneously as an optical source and as a waveguide structure, an expansion mode description of the FEL light permits investigation of the amplification, guiding and coupling efficiency of the individual basis modes to the e-beam. 


In general, the choice of analytical model and expansion basis depends on the FEL geometry and operational characteristics. Since the fields inside the e-beam are optically guided during high-gain, a guided expansion basis is a natural choice, but may be plagued by inefficiency in describing the EM field structure if the guiding characteristics of the e-beam are markedly different from those of the virtual waveguide that yields the particular basis set. (The waveguide is referred to as `virtual' because no such  waveguide is assumed to be present in the physical system.) In Ref \cite{Pinhasi} an analysis is presented for calculating the gain-guided FEL supermode by means of a field expansion using eigenmodes of a hollow waveguide with conducting walls. This method can be useful at short wavelengths (e.g. x-ray FEL), even when no such physical boundary exists and the optical beam is guided only by the e-beam. Clearly, in this example analysis and in general, this approach is legitimate if the presence of the artificial boundary does not affect the physical result. The virtual waveguide dimensions must be taken wide enough so that the fields in the actual problem vanish at the virtual boundaries, but not so wide that many modes are required to describe the radiation field such that the calculation is inefficient or does not converge. As an alternate approach described here, a general method that uses eigenmodes of a smooth, transversely anisotropic dielectric waveguide can be used to model the field while avoiding both the undesirable influence of artificial boundary conditions and precise a priori knowledge of the characteristic transverse dimensions of the waveguide features of the e-beam. This approach also provides flexibility in the form of the expansion basis, which is determined by the refractive index distribution. Of particular interest in any expansion mode approach is the coupling to the naturally occurring Hermite-Gaussian (HG) or Laguerre-Gaussian (LG) modes that describe free-space wave propagation in the paraxial limit. This connection is useful for relating the FEL modes to free-space propagating modes, which are present both before and after the exponential gain process occurs in the undulator. With this connection one may, for example, compactly and clearly describe the input radiaiton needed for efficiently seeding the FEL. One may also robustly describe the propagation of the FEL light after saturation, allowing a clear understanding of the demands made on downstream optics, and on expected mode structure at FEL diagnostics or experiments themselves. In the virtual dielectric waveguide expansion, a form of the refractive index that varies quadratically in the transverse dimension yields a basis of guided LG or HG paraxial waves. This facilitates the desired natural description of the FEL radiation, and is the motivation of the present work. 

Optical guiding in an FEL occurs in the exponential gain regime, when the focusing effects in the source e-beam tend to balance the natural diffraction of the generated EM radiation. Obviously, an expansion set consisting of guided modes can be used to accurately describe the field during long sections of gain-guiding, over which the otherwise free-space fields have many Rayleigh lengths to diffract.  Such an expansion can also accurately describe short sections of dynamic profile evolution (like the FEL start-up period in a high-gain amplifier). But guided modes may not efficiently capture (in the sense of providing a compact description) the field behavior during long sections of weak guiding when diffraction is dominant, like during saturation or low-gain. The regimes of validity for a description using guided modes may be estimated by inspection of the relative values of the optical Rayleigh length $z_R=k r_0^2$ (assuming for the moment that the field has the same characteristic radius as the e-beam, $r_0$), the e-folding radiation power gain length $L_G$, and the overall length of the FEL interaction length $L$. In the guided mode expansion here, we restrict our attention to high-gain ($L_G\ll L$) FEL systems, prior to the onset of saturation, for which the weakly diffracting ($L_G\ll L<z_R$) or strong-guiding ($L_G<z_R<L$) conditions are valid, and the guided mode expansion description can be used efficiently. 

In this work, a virtual dielectric expansion description of high-gain FEL radiation is presented. The excitation of the slowly-growing mode amplitude coefficients in the presence of a local source current is derived. The e-beam is linearized in the cold-beam fluid approximation and a set of coupled excitation equations, modified from Ref \cite{Pinhasi}, is obtained. The coupled equations are then recast by a simple transformation into a matrix equation for solutions of the FEL supermode as a function of parameters for energy detuning, longitudinal space-charge effects and mode coupling to the e-beam. The theory presented here provides a framework for detailed numerical studies performed in a follow-up paper (II. Modeling and numerical simulations). There, results that highlight the advantages of the dielectric eigenmode expansion approach are presented and analyzed.


\section{Dielectric Waveguide Expansion Modes in the presence of a local source}
The radiation fields $\underline{E}(\mathbf{r},t)$ and $\underline{H}(\mathbf{r},t)$ emitted by the currents in the FEL can be expanded in terms of transverse radiation modes $\{\underline{\tilde{\mathcal{E}}}_{\perp},\underline{\tilde{\mathcal{H}}}_{\perp}\}$ of a guiding structure, with slowly-growing amplitudes that vary only as a function of the axis and e-beam propagation coordinate, $z$. Neglecting backward propagating waves (which is well justified in an FEL), the general field expansion is,
\begin{equation}
\begin{aligned}
\underline{\tilde{E}}_{\perp}(\underline{r})&=\sum_q C_{q}(z)\underline{\tilde{\mathcal{E}}}_{\perp q}(r_{\perp})e^{i k_{zq} z} \\ 
\underline{\tilde{H}}_{\perp}(\underline{r})&=\sum_q C_{q}(z)\underline{\tilde{\mathcal{H}}}_{\perp q}(r_{\perp})e^{i k_{zq} z}
\end{aligned}
\label{waveguidemodes}
\end{equation}
where $\underline{\tilde{\mathcal{H}}}_{\perp q}=(1/Z_q)\hat{e}_z\times \underline{\tilde{\mathcal{E}}}_{\perp q}$ is the magnetic field expansion mode, $k_{zq}$ is the $q^{th}$ mode axial wavenumber and the impedance is $Z_q=(k/k_{zq})\sqrt{\mu_0/\epsilon_0}$ for TE modes. In general, the sum extends over both the guided and the cutoff modes, and the modes form a complete set\cite {Marcuse}. The expansion modes are orthogonal and normalized to the mode power
\begin{equation}
\mathcal{P}_q\delta_{q,q'}=\frac{1}{2}\textrm{Re}\Big[\int\!\!\!\int[\underline{\tilde{\mathcal{E}}}_{\perp q}(r_{\perp})\times\underline{\tilde{\mathcal{H}}}^*_{\perp q'}(r_{\perp})]\cdot \hat{e}_z d^2\mathbf{r}_{\perp}\Big].
\label{modepower}
\end{equation}
In free-space, the equation for the full time-harmonic fields in the presence of a source is
\begin{equation}
\nabla^2\underline{\tilde{E}}+k^2\underline{\tilde{E}}=-i\omega\mu_0\underline{\tilde{J}}
\label{fullequation}
\end{equation}
where the total field is $\underline{E}(\mathbf{r},t)=$Re$[\underline{\tilde{E}}(\mathbf{r})e^{-i \omega t}]$ and $k=\omega/c$ is the free-space wavenumber. The transverse charge density gradient of the e-beam in an FEL is typically small and is neglected. For a given choice of expansion basis, the field expansions in Eqs. (\ref{waveguidemodes}) can be inserted into Eq. (\ref{fullequation}) to obtain a differential equation for the amplitude coefficents $C_{q}(z)$\cite{Pinhasi}. Here, we explore the case where the expansion mode $\tilde{\mathcal{E}}_{\perp q}$ is an eigenmode of a dielectric waveguide with refractive index $n(r_{\perp})$. Assuming small transverse variation, $\nabla n(r_{\perp})^2\ll k$, the eigenmodes can be regarded as dominantly transverse, and the dielectric eigenmode equation is
\begin{equation}
\nabla_{\perp}^2\tilde{\underline{\mathcal{E}}}_{\perp q}(r_{\perp})+[n(r_{\perp})^2k^2-k^2_{zq}]\tilde{\underline{\mathcal{E}}}_{\perp q}(r_{\perp})=0.
\label{idealmodes}
\end{equation}
The use of dielectric eigenmodes as an expansion basis must be accompanied by additional terms in Eq. (\ref{fullequation}). These terms are included to offset the virtual polarization currents that arise from the refractive index of the virtual waveguide, since no such waveguide exists along the e-beam axis in the physical system. The polarization current is given by $\underline{\tilde{J}}_p(\underline{r})=i\omega\epsilon_0\chi(r_{\perp})\underline{\tilde{E}}(r_{\perp})=i\omega\epsilon_0(n(r_{\perp})^2-1)\underline{\tilde{E}}(r_{\perp})$ where $n(r_{\perp})=\sqrt{\epsilon(r_{\perp})/\epsilon_0}$. Mathematically, this process amounts to adding $(n(r_{\perp})^2-1)k^2\underline{\tilde{E}}(r_{\perp})$ to both sides of Eq. (\ref{fullequation}). From the dielectric eigenmode equation (\ref{idealmodes}), the modes are assumed to be dominantly transverse, and Eqn (\ref{fullequation}) becomes
\begin{equation}
\nabla^2\underline{\tilde{E}}_{\perp}+n(r_{\perp})^2k^2\underline{\tilde{E}}_{\perp}=-i\omega\mu_0\underline{\tilde{J}}_{\perp}+(n(r_{\perp})^2-1)k^2\underline{\tilde{E}}(r_{\perp}).
\label{dielectricequation}
\end{equation}
Plugging in the expansion fields from Eq. (\ref{waveguidemodes}), the dielectric eigenmode equation in Eq. (\ref{idealmodes}) is used to eliminate the transverse Laplacian term, and the excitation equation for the mode $q$ in the presence of a local source current is given by,
\begin{equation}
\frac{d}{dz}C_q(z)=-\frac{1}{4\mathcal{P}_q}e^{-ik_{zq}z}\int\!\!\!\int\underline{\tilde{J}}_{\perp}(\underline{r})\cdot\tilde{\underline{\mathcal{E}}}^*_{\perp q}(r_{\perp})d^2\mathbf{r}_{\perp}-i\sum_{q'}\kappa^d_{q,q'}C_{q'}(z)e^{-i\Delta k_{zqq'}z}
\label{dielectricexcitation}
\end{equation}
where 
\begin{equation}
\kappa^d_{q,q'}=\frac{\omega\epsilon_0}{4\mathcal{P}_q}\int\!\!\!\int[n(r_{\perp})^2-1]\tilde{\underline{\mathcal{E}}}_{\perp q'}(r_{\perp})\cdot \tilde{\underline{\mathcal{E}}}^*_{\perp q}(r_{\perp})d^2\mathbf{r}_{\perp}.
\label{kappa}
\end{equation}
The difference between the axial wavenumbers of the modes is $\Delta k_{zqq'}=k_{zq}-k_{zq'}$. The term $\kappa^d_{q,q'}$ characterizes the mode overlap in the dielectric, and physically represents the virtual polarization currents that are necessarily subtracted when using eigenmodes of a virtual dielectric waveguide.

\section{electron-beam fluid and coupled excitation equations}
A linear plasma fluid model for a cold e-beam (negligible energy spread) can be used to describe the signal excitation in an FEL interaction\cite{Pinhasi}. A relativistic e-beam in an FEL experiences transverse oscillations, or wiggling, driven by an interaction with a periodic structure or undulator. This periodic motion drives an axial ponderomotive force that modulates the axial electron velocity such that, to first-order, the axial velocity of a cold beam within a static undulator can be expanded as $v_z(\underline{r},t)=v_{z_0}+\mathbf{\textrm{Re}}[\tilde{v}_{z_1}(\underline{r})e^{-i\omega t}]$ where $v_{z_0}=\beta_zc$ is the d.c. component and $\tilde{v}_{z_1}$ is the modulation at signal frequency $\omega$. Longitudinal variations in the velocity like those found in planar undulator systems are ignored for the moment. The velocity modulation $\tilde{v}_{z_1}$ develops a density bunching modulation that is similarly described in a linear model as $\textsf{n}(\underline{r},t)=\textsf{n}_0f(r_{\perp})+\mathbf{\textrm{Re}}[\tilde{\textsf{n}}_1(\underline{r})e^{-i\omega t}]$ where $\textsf{n}_0$ is the on-axis electron density and $f(r_{\perp})$ is the transverse density profile of the e-beam. The lowest order a.c. component of the longitudinal current density results from both the axial velocity and density modulations and is $\tilde{J}_{z}(\underline{r})=-e[\textsf{n}_0f(r_{\perp})\tilde{v}_{z_1}(\underline{r})+v_{z_0}\tilde{\textsf{n}}_1(\underline{r})]$.
If the transverse divergence of the current density modulation is assumed small $\nabla_{\perp}\cdot\tilde{J}_{\perp}\ll\partial\tilde{J}_{z}/\partial z$, the continuity equation can be written as
\begin{equation}
\frac{d}{dz}\tilde{J}_{z}=-i\omega e \tilde{\textsf{n}}_1(\underline{r})
\label{continuity}
\end{equation}

The relativistic force equation for the axial velocity modulation is,
\begin{equation}
\frac{d}{dz}\tilde{v}_{z_1}(\underline{r})-i\frac{\omega}{v_{z_0}}\tilde{v}_{z_1}(\underline{r})=-\frac{e}{\gamma\gamma_{z}^2mv_{z_0}}[\tilde{E}^{SC}_{z}+(\underline{\tilde{v}}\times\underline{\tilde{B}})_z]
\label{relativisticforce}
\end{equation}
where the axial space-charge field $\tilde{E}^{SC}_{z}=\tilde{J}_{z}/i\omega\epsilon_0$ is due to the current density modulation. It is assumed that the e-beam radius is large compared to the bunching wavelength in the e-beam frame, $r_0> \lambda\gamma_z$, so the fringing fields and the transverse space-charge effects are neglected. The interaction between the transverse electron motion and the transverse magnetic fields results in an axial ponderomotive field $(\underline{\tilde{v}}\times\underline{\tilde{B}})_z=\sum_qC_q(z)\tilde{\mathscr{E}}_{pm,q}(r_{\perp})e^{i(k_{zq}+k_w)z}$ where $\tilde{\mathscr{E}}_{pm,q}(r_{\perp})=\frac{1}{2}[\underline{\tilde{v}}_{\perp q}\times\tilde{\underline{\mathcal{B}}}^{*}_{\perp w}+\underline{\tilde{v}}^*_{\perp w}\times\tilde{\underline{\mathcal{B}}}_{\perp q}]\cdot\hat{e}_{z}$, $\tilde{\mathbf{\mathcal{B}}}_{\perp w}$ is the transverse magnetic field of the undulator and $\tilde{v}_{\perp q}(r_{\perp})$ is the transverse electron velocity due to the Lorentz force of the $q^{th}$ mode of the signal field. The undulator wavenumber is identified as $k_w=2\pi/\lambda_w$, and the transverse velocity due to the magnetic undulator field is $\underline{\tilde{v}}_{\perp w}=(-icK/\gamma)\hat{e}_{z}\times\hat{e}_{w}$ where $K=e|\mathcal{\tilde{B}}_{\perp w}|/mck_w$ is the undulator parameter, and $\hat{e}_{w}=\mathcal{\underline{\tilde{B}}}_{\perp w}/|\mathcal{\tilde{B}}_{\perp w}|$ is the unit vector of the undulator field. 
 
By combining Eqs. (\ref{continuity}) and (\ref{relativisticforce}) and using the definition of the modulated longitudinal current density, the density bunching evolution can be expressed as a second order differential equation. We obtain the result from Ref. \cite{Pinhasi} with transverse fields for the density modulation evolution during the FEL interaction:
\begin{equation}
\Big[\frac{d^2}{dz^2}-2i\frac{\omega}{v_{z_0}}\frac{d}{dz}+\frac{\omega^2_{p_0}f(r_{\perp})-\omega^2}{v^2_{z_0}}\Big]\tilde{\textsf{n}}_1(\underline{r})=i\frac{\omega^2_{p_0}f(r_{\perp})}{v^2_{z_0}}\frac{\epsilon_0}{e}\sum_q(k_{zq}+k_w)C_q(z)\tilde{\mathscr{E}}_{pm,q}(r_{\perp})e^{i(k_{zq}+k_w)z}
\label{densitymodulation}
\end{equation}
where the longitudinal relativistic plasma frequency on axis is $\omega^2_{p_0}=e^2\textsf{n}_0/\gamma\gamma^2_z\epsilon_0m$. The beam current is $I_0=-ev_{z_0}\textsf{n}_0\int\!\!\!\int f(r_{\perp})d^2\mathbf{r}_{\perp}$ and the effective beam area through the normalization condition $\int\!\!\!\int f(r_{\perp})d^2\mathbf{r}_{\perp}=A_e$. For a uniform cross-sectional distribution of the e-beam, $f(r_{\perp})=1$ for $r\le r_0$ and zero otherwise. For a gaussian distribution, $f(r_{\perp})=$exp$(-r^2/r_0^2)$.

The transverse component of the current density that excites the signal wave is written in terms of the density modulation as 
\begin{equation}
\underline{\tilde{J}}_{\perp}(\underline{r})=-\frac{1}{2}e\tilde{\textsf{n}}_1(\underline{r})\underline{\tilde{v}}_{\perp w}e^{-ik_wz}.
\label{currentdensity}
\end{equation}
The charge density modulation $\tilde{\textsf{n}}_1(\underline{r})$ appears both in the field mode excitation equation (\ref{dielectricexcitation}) and in the density modulation evolution equation (\ref{densitymodulation}). These equations are coupled through the ponderomotive field, which illustrates the relationship between the density modulation and the excited signal field in an FEL. Both equations can be simplified by expressing the density modulation as a sum over the expansion basis functions and slowly-varying amplitudes, 
\begin{equation}
\tilde{\textsf{n}}_1(\underline{r})=\frac{k\epsilon_0}{e}\sum_q B_{q}(z)\tilde{\mathcal{E}}_{\perp q}(r_{\perp})e^{i\frac{\omega}{v_{z_0}}z}.
\label{densityexpansion}
\end{equation}
Plugging this expression into Eqs. (\ref{currentdensity}), (\ref{densitymodulation}), and (\ref{dielectricexcitation}), and integrating over the transverse dimensions in Eq. (\ref{densitymodulation}), the orthogonality of the eigenfunctions can be used to simplify the equations and to write both coupled equations in terms of the slowly growing amplitudes $C_q(z)$ and $B_q(z)$ of the signal field and the density modulation, respectively. This yields the coupled FEL excitation equations:

\begin{equation}
\begin{aligned}
&\frac{d}{dz}C_q(z)=-i\xi_q\hat{g}_q^*B_q(z) e^{i\theta_qz}-i\sum_{q'}\kappa^d_{q,q'}C_{q'}(z)e^{i(\theta_{q'}-\theta_{q})z}\\
&\frac{d^2}{dz^2}B_q(z)+\theta_{p}^2\sum_j\mathbb{F}_{q,j}B_j(z)=-\frac{1}{\hat{g}_q^*\xi_q}\sum_{q'}Q_{q,q'}C_{q'}(z)e^{-i\theta_{q'} z},
\end{aligned}
\label{coupledSC}
\end{equation}
where $\theta_q=\omega/v_{z_0}-(k_{zq}+k_w)$ is the characteristic detuning parameter for a given mode $q$, $\xi_q=Kk^2/4\gamma k_{zq}$, and $\theta_p=\omega_{p_0}/v_{z_0}$ is the longitudinal plasma wavenumber of a uniformly distributed electron beam profile used in a 1D model. The factor $\hat{g}_q^*=(\hat{e}_z\times\hat{e}_w)\cdot\hat{e}_q^*$ is the polarization alignment factor and measures the relative alignment of the transverse electron motion in the undulator ($\hat{e}_z\times\hat{e}_w$) with the (complex-conjugated) E-field mode polarization direction ($\hat{e}_q^*$). When the wiggling motion direction matches the mode polarization (which may not be the case for the input mode in a seeded FEL scenario), $\hat{g}_q=1$. 

The coupling between the e-beam and the signal field is given by the mode coupling coefficient:
\begin{equation}
Q_{q,q'}=\textrm{JJ}\frac{\theta_p^2\epsilon_0}{8\mathcal{P}_q}(k_{zq'}+k_w)\int\!\!\!\int f(r_{\perp})\tilde{\mathscr{E}}_{pm,q'}(r_{\perp})\underline{\tilde{v}}_{\perp w}\cdot\tilde{\underline{\mathcal{E}}}^*_{\perp q}(r_{\perp})d^2\mathbf{r}_{\perp}
\label{Q}
\end{equation}
where JJ$=[J_0(\alpha)-J_1(\alpha)]^2$ is now included for a strong planar undulator (JJ=1 for a helical undulator geometry). $J_0$ and $J_1$ are the first and second order Bessel functions and $\alpha=K^2/(4+2K^2)$. 
The ponderomotive field $\tilde{\mathscr{E}}_{pm,q'}$ is evaluated explicitly for TE modes in the Appendix, and we obtain a simplified form,
\begin{equation}
Q_{q,q'}=\textrm{JJ}\phantom{l}\theta_p^2\frac{(k_{zq'}+k_w)^2}{8k_{zq}}\Big(\frac{K}{\gamma}\Big)^2\hat{g}_{q}^*\hat{g}_{q'}\mathbb{F}_{q,q'}.
\label{QTEM}
\end{equation}
where $\mathbb{F}_{q,q'}$ is referred to as the beam profile overlap coefficient and quantifies the spatial overlap of the e-beam profile with the expansion modes $q,q'$:
\begin{equation}
\mathbb{F}_{q,q'}=\frac{\int\!\!\!\int f(r_{\perp})\tilde{\mathcal{E}}_{\perp q'}(r_{\perp})\tilde{\mathcal{E}}^*_{\perp q}(r_{\perp})d^2\mathbf{r}_{\perp}}{\int\!\!\!\int|\tilde{\mathcal{E}}_{\perp q}(r_{\perp})|^2d^2\mathbf{r}_{\perp}}.
\label{simpleoverlapfactor}
\end{equation}
This coefficient also appears in connection with the longitudinal plasma wave dynamics (left hand side of the second equation in Eqs. \ref{coupledSC}).

It also may be illuminating to also define the current bunching amplitude:
\begin{equation}
\tilde{\mathfrak{i}}_q(z)=-\frac{1}{4\mathcal{P}_q}e^{-i(\frac{\omega}{v_{z_0}}-k_w)z}\int\!\!\!\int\underline{\tilde{J}}_{\perp}(\underline{r})\cdot\tilde{\underline{\mathcal{E}}}^*_{\perp q}(r_{\perp})d^2\mathbf{r}_{\perp}
\label{bunchingparameter}
\end{equation}
which is interpreted as the slowly-varying amplitude of the transverse optical current. Replacing $\underline{\tilde{J}}_{\perp}$ with it's definition from Eq. (\ref{currentdensity}) and plugging in the density modulation expansion from Eq. (\ref{densityexpansion}), the integration over the transverse space is straight-forward by the orthogonality of the eigenmodes. Thus, a simplified expression is obtained for $\tilde{\mathfrak{i}}_q(z)$ in terms of the density modulation amplitudes: 
\begin{equation}
\tilde{\mathfrak{i}}_q(z)=-i\xi_q\hat{g}_q^*B_q(z). 
\label{bunchingparameter1}
\end{equation}
Substitution of this into Eq. (\ref{coupledSC}) allows one to solve the coupled FEL excitation equations in terms of the optical current amplitudes.

In the 1D limit, the axial wavenumbers are degenerate $k_{zq}=k$, the e-beam profile is constant, $f(r_{\perp})\rightarrow1$, and the overlap factor becomes $\mathbb{F}_{q,q'}=\delta_{q,q'}$. The e-beam coupling coefficient is then written in 1D simplified form, $Q=\textrm{JJ}k(\theta_pK/2\sqrt{2}\gamma\beta_z)^2=(2k_w\rho)^3$, where $\rho=(\textrm{JJ}e^2K^2\textsf{n}_0/32\epsilon_0\gamma^3mc^2k_w^2)^{1/3}$ is the well-known Pierce parameter often used in FEL theory.

The first of the coupled equations in Eqs. (\ref{coupledSC}) describes the excitation of the mode amplitude $C_q$ of a dielectric waveguide eigenmode due to the transverse wiggling motion of the bunching current throughout the FEL interaction. The second equation in (\ref{coupledSC}) describes the evolution of the density bunching amplitude, which is excited in the e-beam by the EM signal field. The effect of the longitudinal space-charge in the beam is calculated in the second term of the second equation, and takes into account the effects of longitudinal plasma oscillations (Langmuir waves) in the FEL interaction. The effects of fringing fields due to a transverse variation in the axial space-charge field can be examined in a more complete 3D scenario\cite{Steinberg}, but are presently neglected in the approximation that $r_0> \lambda\gamma_z$. 

The initial conditions for Eqs. (\ref{coupledSC}) specify the operating characteristic of the FEL.  For example, when operating as a single-pass amplifier (seeded FEL) there is negligible initial density and velocity modulation ($B_{q}(0),dB_{q}(z)/dz|_{z=0}=0$) and the initial seed field is non-zero ($C_{q}(0)\ne0$). For a self-amplified spontaneous emission (SASE) FEL, the amplified shot noise can be related to the pre-bunching conditions ($B_{q}(0)\ne0,dB_{q}(z)/dz|_{z=0}\ne0$) and the input signal field vanishes ($C_{q}(0)=0$). 

The coupled expressions in Eqs. (\ref{coupledSC}) describe the evolution of the e-beam modulation currents and the excitation of the EM signal field during the FEL interaction, inclusive of longitudinal space-charge effects. During high-gain operations when the radiation field grows exponentially, the self-guiding effects become dominant over the diffraction, and the radiation field can be accurately described by a collection of waveguide modes. The virtual dielectric eigenmode basis used for the radiation field expansion can be any complete set that satisfies the dielectric waveguide equation (\ref{idealmodes}). 
In the general expansion presented here, the form for $n(r_{\perp})$ can be freely chosen to yield different functional forms for the expansion modes, allowing flexibility to choose a basis that is perhaps better suited to describe a given FEL system. Some of the most typical forms include an index with a quadratic spatial dependence which results in Hermite-Gaussian or Laguerre-Gaussian expansion modes\cite{YarivQE} (examined in detail for LG modes in part II of this work), or a step-profile optical fiber\cite{Marcuse,Scharlemann} which yields Bessel functions. The closer that the choice of the virtual dielectric waveguide distribution is to the real solution to the given FEL system, the fewer the number of modes that will be required to converge the equations in Eqs. (\ref{coupledSC}) to the correct solution for the FEL.

\section{Supermode Matrix Solutions}
In the high-gain regime of an FEL with finite e-beam width, the radiation field tends to concentrate near the beam and produce a power amplified radiation wave with a self-similar transverse field distribution that propagates along the interaction length. This specific complex-valued combination of the expansion modes is referred to as the supermode, or the eigenmode of the high-gain FEL. In a long undulator, the fundamental supermode evolves spontaneously since it has the highest gain. The initial conditions (i.e. the transverse profile and phase distribution of the injected radiation field or density pre-bunching) affect the mode excitation composition and evolution. However, these initial conditions only affect the supermode establishment length. Evenutally, in an undepleted system, the supermode will become dominant in a long enough undulator, prior to the onset of saturation. 

To find the optimal injection parameters that match and expedite the establishment of the supermode, one must solve Eqs. (\ref{coupledSC}) with the appropriate initial conditions. To simply find the system supermodes however, it is enough to find the eigensolutions to Eqs. (\ref{coupledSC}). These are the combinations of the expansion mode profiles that propagate self-similarly, i.e., with constant amplitude coefficients and with distinct complex wavenumbers\cite{Saldin}. In the presence of gain, each supermode wavenumber will be different from the wavenumber of free-space and can be written with a perturbative term $\tilde{\delta k}$ that is due to the FEL interaction:
\begin{equation}
k_{SM}=k+\tilde{\delta k},
\label{SMwavenumber}
\end{equation}
where $\textrm{Re}\{\tilde{\delta k}\}$ anticipates an effective modified refractive index to that of free-space, and $\textrm{Im}\{\tilde{\delta k}\}$ is related to the gain. Since the supermodes evolve after the initial startup period and have fixed transverse profiles along $z$, one can substitute $C_q(z)=b_qe^{i(k_{SM}-k_{zq})z}$ for the mode amplitude coefficients Eq. (\ref{waveguidemodes}). The mode amplitude coefficients $b_q$ are constants, and the $z$-dependence is contained solely in the mode-independent exponential term. The time-harmonic electric field of the supermode is then
\begin{equation}
\underline{\tilde{E}}_{SM}(\underline{r})=\Big[\sum_q b_q\tilde{\underline{\mathcal{E}}}_{q}(r_{\perp})\Big]e^{ik_{SM}z}.
\label{supermodes}
\end{equation}
This describes a transverse field that is fixed in transverse profile, but is growing exponentially in amplitude along $z$. Inserting this transformation into Eq. (\ref{coupledSC}) converts the coupled second-order differential equations into a set of coupled algebraic equations:
\begin{equation}
(\tilde{\delta k}-\theta)^2\Big[(\tilde{\delta k}-\Delta k_{q})b_q+\sum_{q'}\kappa^d_{q,q'}b_{q'}\Big]-\theta_p^2\sum_j\Bigg{\{}\frac{k_{zj}}{k_{zq}}\mathbb{F}_{q,j}\Big[(\tilde{\delta k}-\Delta k_{j})b_j+\sum_{q''}\kappa^d_{j,q''}b_{q''}\Big]\Bigg{\}}+\sum_{q'''}Q_{q,q'''}b_{q'''}=0,
\label{smlinear}
\end{equation}
where $\theta=\omega/v_{z_0}-k-k_w$ is the detuning parameter for a 1D model ($k_{zq}=k$), and $\Delta k_{q}=k_{zq}-k$. The coupled equations in (\ref{smlinear}) can be solved to yield values for the supermode coefficients $b_q$ in terms of the perturbation $\tilde{\delta k}$. It is convenient to write Eqs (\ref{smlinear}) in a simplified matrix determinant form
\begin{equation}
\Bigg{|}\Big[\underline{\underline{\mathbf{I}}}(\tilde{\delta k}-\theta)^2-\theta_p^2\phantom{l}\underline{\underline{\mathbb{M}}}\Big]\Big[\underline{\underline{\mathbf{I}}}\tilde{\delta k}+\underline{\underline{\kappa^d}}-\underline{\underline{\Delta k}}\Big]+\underline{\underline{Q}}\Bigg{|}=0.
\label{supermodedeterminant}
\end{equation}
The matrix elements of $\underline{\underline{\mathbb{M}}}$ are given by $\mathbb{M}_{q,q'}=(k_{zq'}/k_{zq})\mathbb{F}_{q,q'}$, and similarly for $\underline{\underline{\kappa^d}}=\{\kappa^d_{q,q'}\}$, $\underline{\underline{Q}}=\{Q_{q,q'}\}$, and $\underline{\underline{\Delta k}}=\{\Delta k_{q}\delta_{q,q'}\}$. The matrix $\underline{\underline{\mathbf{I}}}$ is the identity. 

The solutions to Eq. (\ref{supermodedeterminant}) yield $3N$ solutions for $\tilde{\delta k}$, where $N$ is the number of expansion modes. Each $\tilde{\delta k}$ can then be inserted in Eq. (\ref{smlinear}) to find a non-trivial solution (if one exists) for the mode amplitude coefficients of an eigenmode of the FEL system. From Eq. (\ref{supermodes}) and the definition of $k_{SM}$ in Eq. (\ref{SMwavenumber}), it can be seen that the solution for $\tilde{\delta k}$ with the most negative imaginary component drives the highest gain, and dominates over the rest of the eigenmodes. This value is used in solving Eq. (\ref{smlinear}) and will yield the coefficients of the dominant supermode, with the 3D power gain length, or e-folding length, given by $L_G=(2|$Im$\{\tilde{\delta{k}}\}|)^{-1}$.

We note that when $\underline{\underline{\Delta k}},\underline{\underline{\kappa^d}}=0$, the matrix equation in (\ref{supermodedeterminant}) reduces to a generalized matrix form of the canonical FEL cubic equation with longitudinal space-charge. In the additional limit of a large, effectively constant transverse beam profile, $f(r_{\perp})\rightarrow$ 1 so $\mathbb{F}_{q,q'}=\delta_{q,q'}$ and $Q_{q,q'}\rightarrow Q\delta_{q,q'}$, and Eq.  (\ref{supermodedeterminant}) takes the form of the familiar 1D FEL cubic equation with the mode-independent beam coupling parameter $Q$ and the axial plasma wavenumber $\theta_p$.

\section{Single Guided Gaussian Mode Approximation}
During high-gain, the proper balance between the natural diffraction of the coherent radiation and the guided focusing of the radiation due to the e-beam determines the eventual spot size $w_{SM}$ of the EM supermode field. This can be obtained with the dielectric expansion formulation in a natural way, through the solutions to the excitation equations (\ref{coupledSC}) or the supermode determinant equation (\ref{supermodedeterminant}), using $N$ modes in the field expansion. Clearly, choosing a suitable expansion basis (i.e., one that is close in form to that of the supermode field) can greatly reduce the number of expansion modes required to precisely describe the FEL supermode. In some cases, it may even be sufficient to use only a single mode to model the field, in order to estimate the relevant supermode characteristics without solving the full equations. A single mode approximation can also be used to streamline the computation of the full solutions by providing an approximate value for the expansion basis waist size $w_0$ which is used as a scaling parameter in the basis eigenfunctions $\tilde{\mathcal{E}}_{q}(r_{\perp};w_0)$. The supermode waist size (and thus the parameter $w_0$) can sometimes be roughly approximated as equal to the e-beam radius $w_{SM}\sim r_0$ for an axisymmetric e-beam distribution, but the actual value may vary significantly depending on the operating parameters of the FEL. To better quantify this estimate for the general case, we analyze the guided supermode spot size by approximating it as a single gaussian mode (SGM) that propagates with a fixed spot size, $w_{g}$. Gaussian modes, like the HG and LG modes discussed previously, are found as solutions to Eq. (\ref{idealmodes}) with a refractive index $n(r)^2$ that varies quadratically in the transverse dimension, describing what is known as a weakly-guiding \emph{quadratic index medium}, or QIM. In this context, the SGM analysis also provides insight into the physical origin of the characteristic guiding of the supermode in terms of guided paraxial gaussian modes from conventional fiber waveguide theory. 

The refractive index that yields axisymmetric guided gaussian eigenmodes in Eq. (\ref{idealmodes}) is 
\begin{equation}
n^2(r)=n^2_0-\Big(\frac{r}{z_R}\Big)^2
\label{QIM}
\end{equation}
where $z_R=kw_g^2/2$ is the Rayleigh length and $w_g$ is the characteristic rms radius of the gaussian field profile given by $\tilde{\mathcal{E}}_{q}(r_{\perp};w_g)$. The axial wavenumber, in the paraxial approximation is therefore given by
\begin{equation}
k_{z0}\simeq kn_0-\frac{2}{n_0kw_{g}^2}.
\label{gaussianwavenumber}
\end{equation}
A single gaussian mode then corresponds to only the $(q,q')=(0,0)$ matrix elements in the supermode determinant in Eq. (\ref{supermodedeterminant}). For $n_0\simeq1$, an explicit analytic expression is obtained for the dielectric mode coupling parameter $\kappa^d_{0,0}=-1/kw_{g}^2$. Ignoring the effects of space-charge waves ($\theta_{p}L<\pi$), the form of Eq. (\ref{supermodedeterminant}) for a single gaussian mode is
\begin{equation}
(\tilde{\delta k}-\theta)^2\big[\tilde{\delta k}+\frac{1}{k w_{g}^2}\big]+Q_{0,0}=0.
\label{singlemodematrix}
\end{equation}
where $\Delta k_0\simeq-2/kw_{g}^2$ has been obtained from Eq. (\ref{gaussianwavenumber}). 
Equation (\ref{singlemodematrix}) then becomes a modified 1D FEL cubic equation where $Q_{0,0}=Q\phantom{l}\mathbb{F}_{0,0}$ is the SGM gain parameter, which describes the modification of the 1D e-beam mode coupling parameter $Q$ by the filling factor (Eq. \ref{simpleoverlapfactor}) of the fundamental gaussian EM mode, $\mathbb{F}_{0,0}$.
One can then define a shifted perturbation parameter:
\begin{equation}
\delta k=\tilde{\delta k}+\frac{1}{k w_{g}^2}
\label{paraxialperturbation}
\end{equation}
and then set the detuning to $\theta=-1/k w_{g}^2$. Equation (\ref{singlemodematrix}) then becomes a simple 1D FEL cubic equation at resonance,
\begin{equation}
\delta k^3+Q_{0,0}=0
\label{singlemodecubic}
\end{equation}
In this form, the solution for $\delta k$ that corresponds to the dominant high-gain mode is straightforward, and well-known from FEL theory:
\begin{equation}
\delta k=\frac{1-i\sqrt{3}}{2}Q_{0,0}^{1/3}.
\label{deltak}
\end{equation}

In a high-gain FEL, the supermode wavenumber $k_{SM}$ in Eq. (\ref{SMwavenumber}) is defined as the wavenumber of a plane wave $k$ that is modified through the FEL interaction by the perturbative factor $\tilde{\delta k}$. Combining the definition of $k_{SM}$ in Eq. (\ref{SMwavenumber}) with $\delta k$ from Eq. (\ref{paraxialperturbation}) we can write
\begin{equation} 
k_{SM}=k+ \delta k-\frac{1}{k w_{g}^2}
\label{SGM1}
\end{equation}
for a single gaussian supermode. Since the supermode is assumed to propagate as a guided gaussian mode with a fixed spot size and velocity, the real (propagating) part of the supermode wavenumber is equated to the propagating gaussian mode wavenumber in the dielectric waveguide:
\begin{equation}
\textrm{Re}\{k_{SM}\}=k_{z0}.
\label{SGM}
\end{equation}
Combining the real parts of Eqs. (\ref{deltak}), (\ref{SGM1}) and (\ref{SGM}), and recognizing $n_0=1+Q^{1/3}/2k\simeq 1$ as the effective refractive index on axis from a 1D FEL model, we obtain a relation for the supermode spot size in terms of the 1D coupling and the filling factor
\begin{equation}
\frac{1}{k w_{g}^2}=\frac{Q^{1/3}}{2}\big(1-\mathbb{F}_{0,0}^{1/3}).
\end{equation}
Assuming that the e-beam has a gaussian transverse profile $f(r_{\perp})=$exp$(-r^2/r_0^2)$, the filling factor is simply $\mathbb{F}_{0,0}=(1+w_{g}^2/2r_0^2)^{-1}$. We finally obtain an expression for the supermode spot size $w_{g}$ in terms of the 1D coupling parameter $Q$ and the e-beam radius $r_0$:
\begin{equation}
(1+\frac{w_{g}^2}{2r_0^2})\Big[1-\frac{2}{k w_{g}^2Q^{1/3}}\Big]^3=1
\label{singlemode}
\end{equation}
Solutions to Eq. (\ref{singlemode}) for $w_g$ can be easily obtained numerically for given FEL running parameters. The calculated SGM spot size provides a useful prediction of the supermode spot size $w_{SM}$ for a gaussian e-beam lasing at the fundamental and which in turn suggests an appropriate value for the expansion waist $w_0$ to facilitate computational efficiency for solving the full excitation equations.

It is useful to note that while $k_{SM}$ was defined as a modification of a plane wave in Eq. (\ref{SMwavenumber}), in the SGM approximation, one could also consider defining $k_{SM}$ through a modification of a \emph{free-space} paraxial gaussian mode. In other words, to consider the effect of the FEL interaction on converting a free-space gaussian wave to a guided gaussian mode of the FEL.

Free-space modes evaluated at the optical beam waist $w_0$ have an associated axial wavenumber 
\begin{equation}
\hat{k}^2_{z0}=k^2-\frac{2}{w^2_0}.
\label{freespacegaussian}
\end{equation}
Expressing the supermode wavenumber as a modified paraxial wave of free-space, it takes the form
\begin{equation}
k_{SM}=\hat{k}_{z0}+\delta k
\label{singlemodeparaxial}
\end{equation} 
where, just as before, the supermode is a guided eigenmode of the QIM, $\textrm{Re}\{k_{SM}\}=k_{z0}$, but it is now also expressed as a deviation from the gaussian free-space mode $\hat{k}_{z0}$ by the FEL interaction given by $\delta k$. Equating the real parts from Eqs. (\ref{SMwavenumber}) and (\ref{singlemodeparaxial}), we find that $k+\textrm{Re}\{\tilde{\delta k}\}=\hat{k}_{z0}+\textrm{Re}\{\delta k\}$. By rearranging and using $\hat{k}_{z0}-k\simeq-1/kw^2_{g}$ for a single free-space mode, we recover the exact expression for $\delta k$ in Eq. (\ref{paraxialperturbation}). Thus, the SGM approximation of an FEL with a gaussian transverse e-beam profile describes the conversion of plane wave into a guided gaussian mode at a detuning of $\theta=-1/k w_{g}^2$ in an equivalent way as it describes the conversion of a free-space paraxial gaussian mode into a guided gaussian mode, operating at resonance ($\theta=0$).

\section{Conclusions}
The signal field of a high-gain FEL has been described in terms of a sum over orthogonal eigenmodes of a virtual dielectric waveguide. A set of coupled excitation equations for the e-beam density modulations and the field amplitudes in the presence of longitudinal space-charge effects has been derived. This approach can be effectively used to predict relevant FEL parameters such as signal field spot size and intensity distribution for any point along the interaction length, and for any arbitrary initial conditions. It further provides a novel and practical method to analyze the coupling and generation of particular modes, such as those associated with free-space propagation, in the FEL system. It is a general method for calculating the optically guided FEL eigenmodes for any transverse e-beam current and profile distribution, based on a straightforward numerical solution to an algebraic dispersion equation. A follow-up analysis of this work will explore these concepts in detail.

\section{Acknowledgments}
The authors would like to thank Sven Reiche for helpful discussions. This research was supported by grants from Department of Energy Basic Energy Science contract DOE DE-FG02-07ER46272 and Office of Naval Research contract ONR N00014-06-1-0925.

\appendix
\section{e-beam coupling coefficient}
\label{app:1}
A generalized, explicit expression for the mode coupling coefficient $Q_{q,q'}$ of any spatial expansion basis can be obtained starting from the ponderomotive field
\begin{equation}
\tilde{\mathscr{E}}_{pm,q}(r_{\perp})=\frac{1}{2}[\underline{\tilde{v}}_{\perp q}\times\tilde{\underline{\mathcal{B}}}^*_{\perp w}+\underline{\tilde{v}}^*_{\perp w}\times\tilde{\underline{\mathcal{B}}}_{\perp q}]\cdot\hat{e}_{z}.
\label{pm1:app}
\end{equation}
From the relativistic Lorentz force equation we can write the transverse velocities as\cite{Steinberg},
\begin{equation}
\begin{aligned}
&\underline{\tilde{v}}_{\perp w}=-i\frac{e}{\gamma m k_w}\hat{e}_z\times\tilde{\underline{\mathcal{B}}}_w\\
&\underline{\tilde{v}}_{\perp q}(r_{\perp})=-i\frac{e}{\gamma m (\omega-k_{zq}v_{z_0})}[\tilde{\underline{\mathcal{E}}}_{\perp q}(r_{\perp})
+v_{z_0}\hat{e}_z\times\tilde{\underline{\mathcal{B}}}_{\perp q}(r_{\perp})]
=-i\frac{e}{\gamma m \omega}\tilde{\underline{\mathcal{E}}}_{\perp q}(r_{\perp})
\end{aligned}
\label{transversevelocity:app}
\end{equation}
where in the last equality for $\underline{\tilde{v}}_{\perp q}$ it has been assumed that $\tilde{\underline{\mathcal{E}}}_{\perp q}=-(\omega/k_{zq})\hat{e}_z\times\tilde{\underline{\mathcal{B}}}_{\perp q}$ for TE modes, in keeping with the dominantly TEM mode expansion sets of the present work. With the equations in (\ref{transversevelocity:app}) the expression for $\tilde{\mathscr{E}}_{pm,q}$ becomes
\begin{equation}
\tilde{\mathscr{E}}_{pm,q}(r_{\perp})=i\frac{e(k_{zq}+k_w)}{2\gamma m\omega k_w}(\hat{e}_z\times\tilde{\underline{\mathcal{B}}}_{w})^*\cdot\tilde{\underline{\mathcal{E}}}_{\perp q}(r_{\perp}).
\label{pm2:app}
\end{equation}

The complex phasor notation for the undulator field $\underline{\mathcal{B}}_{w}=$Re$\{\tilde{\underline{\mathcal{B}}}_{w}e^{-ik_wz}\}$ allows a general description for any type of undulator polarization. Defining a general transverse field polarization unit vector as $\hat{e}_w=\tilde{\underline{\mathcal{B}}}_{w}/|\tilde{\mathcal{B}}_{w}|$, for a linear undulator $\hat{e}_w=\hat{e}_y$ (with vertical polarization), the field is given as
\begin{equation}
\underline{\mathcal{B}}_{w}=|\tilde{\mathcal{B}}_{w}|\hat{e}_y\cos{k_wz}
\end{equation}
and in a helical undulator $\hat{e}_w=(\hat{e}_x\pm i\hat{e}_y)/\sqrt{2}$ for right-handed (+) or left-handed (-) field orientations and
\begin{equation}
\underline{\mathcal{B}}_{w}=\frac{|\tilde{\mathcal{B}}_{w}|}{\sqrt{2}}(\hat{e}_x\cos{k_wz}\pm \hat{e}_y\sin{k_wz}).
\end{equation}
The undulator parameter is defined as $K=e|\tilde{\mathcal{B}}_{w}|/(k_wmc)$. Note that the same $K$ value corresponds to an on-axis maximum field amplitude of $|\tilde{\mathcal{B}}_{w}|$ in a linear undulator, and an on-axis maximum field amplitude of $|\tilde{\mathcal{B}}_{w}|/\sqrt{2}$ in a helical undulator. In terms of $K$, Eq. (\ref{pm2:app}) can be written as
\begin{equation}
\tilde{\mathscr{E}}_{pm,q}(r_{\perp})=i\frac{k_{zq}+k_w}{2k}\frac{K}{\gamma}(\hat{e}_z\times\hat{e}^*_w)\cdot\tilde{\underline{\mathcal{E}}}_{\perp q}(r_{\perp}).
\label{pm3:app}
\end{equation}
Substitution of this expression and the mode power normalization from Eq. (\ref{modepower}) into $Q_{q,q'}$ results in a general expression for the mode coupling coefficents:
\begin{equation}
Q_{q,q'}=\textrm{JJ}\phantom{l}\theta_p^2\frac{(k_{zq'}+k_w)^2}{8k_{zq}}\Big(\frac{K}{\gamma}\Big)^2\hat{g}_{q}^*\hat{g}_{q'}\mathbb{F}_{q,q'}
\label{kappa:app}
\end{equation}
where $\hat{g}_{q}=(\hat{e}_z\times\hat{e}_w^*)\cdot\hat{e}_q$ is the polarization alignment factor, which measures the relative direction of transverse electron motion in the undulator compared with the electric field polarization of the mode $q$. The spatial overlap factor $\mathbb{F}_{q,q'}$ is defined as
\begin{equation}
\mathbb{F}_{q,q'}=\frac{\int\!\!\!\int f(r_{\perp})\tilde{\mathcal{E}}_{\perp q'}(r_{\perp})\tilde{\mathcal{E}}^*_{\perp q}(r_{\perp})d^2\mathbf{r}_{\perp}}{\int\!\!\!\int|\tilde{\mathcal{E}}_{\perp q}(r_{\perp})|^2d^2\mathbf{r}_{\perp}}.
\label{fulloverlapfactor:app}
\end{equation}
The general expression for the mode coupling coefficient in Eq. (\ref{kappa:app}) permits analysis of mode coupling for radiation modes that vary transversely in their complex field amplitude and in their polarization relative to the undulator. It is also clear that in order to obtain maximal coupling between two modes $q$ and $q'$, they must be polarization matched to each other ($\hat{g}_{q}=\hat{g}_{q'}$) and to the direction of electron motion in the undulator ($\hat{g}_{q}=1$), as well as have finite spatial overlap with both the e-beam distribution. In the single mode limit ($q=q'$), Eq. (\ref{fulloverlapfactor:app}) simplifies to the commonly used ``filling factor" parameter.
\bibliography{DielectricExpansion}

\end{document}